\documentclass[twocolumn,trackchanges]{aastex631}

\newcommand\aastex{AAS\TeX}
\newcommand\latex{La\TeX}

\begin{document}

\title{Production of radioactive $^{22}$Na in core-collapse supernovae: the Ne-E(L) component in presolar grains and its possible consequences on supernova observations}

\author[0000-0002-9048-6010]{Marco Pignatari} 
\affiliation{HUN REN Konkoly Observatory, CSFK, H-1121, Budapest, Konkoly Thege M. \'ut 15–17, Hungary}
\affiliation{CSFK, MTA Centre of Excellence, Budapest, Konkoly Thege Miklós út 15-17., H-1121, Hungary}
\affiliation{University of Bayreuth, BGI, Universitätsstraße 30, 95447 Bayreuth, Germany}
\affiliation{NuGrid Collaboration, \url{http://nugridstars.org}}

\author[0000-0003-4899-0974]{Sachiko Amari}
\affiliation{McDonnell Center for the Space Sciences \& Physics Department, Washington University, St. Louis, MO 63130}

\author{Peter Hoppe}
\affiliation{Max Planck Institute for Chemistry, Hahn-Meitner-Weg 1, D-55128 Mainz, Germany}

 
\author[0000-0003-2624-0056]{C. Fryer}
\affiliation{Center for Theoretical Astrophysics, Los Alamos National Laboratory, Los Alamos, NM 87545, USA}
\affiliation{Center for Nonlinear Studies, Los Alamos National Laboratory, Los Alamos, NM 87545, USA}

\author[0000-0003-3970-1843]{S. Jones}
\affiliation{Theoretical Division, Los Alamos National Laboratory, Los Alamos, NM 87545, USA}

\author[0000-0003-2197-0797]{A. Psaltis}
\affiliation{Department of Physics, North Carolina State University, Raleigh, NC 27695, USA}
\affiliation{Triangle Universities Nuclear Laboratory, Duke University, Durham, NC 27710, USA}

\author{A.M. Laird}
\affiliation{School of Physics, Engineering and Technology, University of York, Heslington, York, YO10 5DD, United Kingdom}

\author[0000-0001-8087-9278]{F. Herwig}
\affiliation{Department of Physics \& Astronomy, University of Victoria, Victoria, B.C. V8W 2Y2, Canada}

\author[0000-0003-0390-8770]{L. Roberti}
\affiliation{Istituto Nazionale di Fisica Nucleare -Laboratori Nazionali del Sud, Via Santa Sofia 62, 95123 CATANIA (Italy)}
\affiliation{HUN REN Konkoly Observatory, Research Centre for Astronomy and Earth Sciences (CSFK), H-1121, Budapest, Konkoly Thege M. \'ut 15–17, Hungary}
\affiliation{CSFK, MTA Centre of Excellence, Budapest, Konkoly Thege Miklós út 15-17., H-1121, Hungary}
\affiliation{NuGrid Collaboration, \url{http://nugridstars.org}}

\author[0000-0002-0552-3535]{Thomas Siegert}
\affiliation{Julius-Maximilians-Universit \"{a}t W \"{u}rzburg, Fakult \"{a}t f\"{u}r Physik und Astronomie, Institut f\"{u}r Theoretische Physik und Astrophysik, Lehrstuhl f\"{u}r Astronomie, Emil-Fischer-Str 31, D-97074 W\"{u}rzburg, Germany}

\author{Maria Lugaro}
\affiliation{HUN REN Konkoly Observatory, Research Centre for Astronomy and Earth Sciences (CSFK), H-1121, Budapest, Konkoly Thege M. \'ut 15–17, Hungary}
\affiliation{CSFK, MTA Centre of Excellence, Budapest, Konkoly Thege Miklós út 15-17., H-1121, Hungary}
\affiliation{ELTE E\"{o}tv\"{o}s Lor\'and University, Institute of Physics and Astronomy, Budapest 1117, P\'azm\'any P\'eter s\'et\'any 1/A, Hungary}
\affiliation{School of Physics and Astronomy, Monash University, VIC 3800, Australia}



\begin{abstract}
Presolar graphite grains carry the isotopic signatures of their parent stars. 
A significant fraction of presolar graphites shows isotopic abundance anomalies relative to solar for elements such as O, Si, Mg and Ca, which are compatible with nucleosynthesis in core-collapse supernovae (CCSNe). Therefore, they must have condensed from CCSN ejecta  before the formation of the Sun. 
Their most puzzling abundance signature is the $^{22}$Ne-enriched component Ne-E(L), interpreted as the effect of the radioactive decay of $^{22}$Na ($T_{1/2}$ = 2.6 years). Previous works have shown that if H is ingested into the He shell and not fully destroyed before the explosion, the CCSN shock in the He shell material produces large $^{22}$Na amounts. 
Here we focus on such CCSN models, showing a radioactive $^{26}$Al production compatible with grains measurements, and analyze the conditions of $^{22}$Na nucleosynthesis. In these models, $^{22}$Na is mostly made in the He shell, with a total ejected mass varying 
between 2.6$\times$10$^{-3}$ M$_{\odot}$ and 1.9$\times$10$^{-6}$ M$_{\odot}$. We show that such $^{22}$Na may already impact the CCSN light curve 500 days after the explosion, and at later stages it can be the main source powering the CCSN light curve for up to a few years before the $^{44}$Ti decay becomes dominant. Based on the CCSN yields above, the 1274.53 keV $\gamma$-ray flux due to $^{22}$Na decay could be observable for years after the first CCSN light is detected, depending on the distance. This makes CCSNe possible sites to detect a $^{22}$Na $\gamma$-ray signature consistently with the Ne-E(L) component found in presolar graphites. Finally, we discuss the potential contribution from the $^{22}$Na decay to the galactic positron annihilation rate.
\end{abstract}


\section{Introduction}
\label{sec:intro}

The radioactive isotope $^{22}$Na ($T$$_{1/2}$ = 2.6 years) can be made 
by proton-capture reactions and the $\beta$$^+$ decay of $^{22}$Mg in different stellar environments, such as classical novae \citep[e.g.,][]{jose:98} and core-collapse supernovae \citep[CCSNe, e.g.,][]{woosley:95}. $^{22}$Na decays to the stable isotope $^{22}$Ne relatively shortly after its production, with its half-life orders of magnitude smaller than typical stellar and astronomical timescales. 

However, the $^{22}$Na alive at the time of the ejection, and now decayed, is measurable today thanks 
to meteoritic stardust analysis. \cite{black:69} first observed the so-called Ne-E(L) component in carbonaceous meteorites, which is characterized by an extremely high $^{22}$Ne enrichment relative to the solar abundance. This was later identified as being carried by presolar graphite grains 
\citep[][]{amari:90}. 
Instead, much smaller or negligible signatures were detected for the other two Ne stable isotopes $^{20}$Ne and $^{21}$Ne\footnote{The most abundant Ne stable isotope in nature is $^{20}$Ne, with the solar Ne composition being 90.48\%, 0.27\% and 9.25\% for $^{20}$Ne, $^{21}$Ne and $^{22}$Ne, respectively (\url{https://www.nndc.bnl.gov/nudat3/}).}. Since \cite{clayton:75}, the commonly-accepted explanation of the $^{22}$Ne-rich Ne-E(L) component is that the observed $^{22}$Ne condensed in grains as $^{22}$Na. Being a noble gas, the original stellar Ne 
could not condense in dust, and 
its component directly implanted in the formed grains seems to be extremely small. 
Therefore, the $^{22}$Ne excess derives from condensation of $^{22}$Na into the grains. 
As a confirmation of this scenario, \cite{heck:18} analysis of low-density graphite grains from the Orgueil (CI1) meteorite found that ($^{22}$Na-) $^{22}$Ne-rich grains are quite common: 6 out of 7 grains on one mount are $^{22}$Ne-rich and 5 of them carry an isotopic signature compatible with a CCSN origin (the remaining $^{22}$Ne-rich grain, 8m-13, was proposed to be formed instead by a low-mass, low-metallicity AGB star). If $^{22}$Ne would be implanted, other Ne isotopes, $^{20}$Ne and $^{21}$Ne, should also have been implanted. In Murchison low-density grains, only
$^{22}$Ne was detected instead \citep[][]{nichols:94}. 
Of the 5 $^{22}$Ne-rich Orgueil low-density grains from CCSNe mentioned above, only one grain contained measurable $^{21}$Ne \citep[][]{heck:18}.
For this grain with $^{21}$Ne, it was estimated that the nucleosynthetic $^{22}$Ne (i.e., implanted $^{22}$Ne) would be 0.5\% of the total $^{22}$Ne in the grain. Therefore, even in this peculiar case, the majority of the $^{22}$Ne measured is still from $^{22}$Na.

Multiple isotopic anomalies measured in single graphites indicate that a significant fraction of them condensed from CCSN ejecta \citep[e.g.,][]{travaglio:99}. Among others, the measured range of C isotopic ratio and excesses in $^{15}$N, $^{26}$Al, $^{28}$Si and $^{44}$Ti are all signatures shared with presolar SiC grains of type X, also made in CCSN ejecta \citep[][and references therein]{zinner:14, nittler:16, boujibar:21, liu:24a}. However, the $^{22}$Ne-rich Ne-E(L) component is still not quantitatively explained by CCSN simulations.

\cite{amari:09} discussed the possibility to reproduce the measured upper limit of 0.01 for $^{20}$Ne/$^{22}$Ne in single graphites \citep[][]{nichols:92}, using predictions from CCSN models by \cite{rauscher:02} and \cite{chieffi:04}. Even the lowest $^{20}$Ne/$^{22}$Ne ratio in CCSNe found in the He/C zone was almost an order of magnitude higher than the upper limit of 0.01 (0.096 and 0.088 for \cite{rauscher:02} and \cite{chieffi:04}, respectively) and no ejecta matched the observed Ne isotope ratios. Therefore, \cite{amari:09} confirmed that Ne-E(L) was not generated by Ne isotopes directly implanted in dust, supporting the $^{22}$Na cosmogenic origin. The former O-rich C shell \citep[or the O/Ne zone,][]{meyer:95} is the most $^{22}$Na-rich part of typical CCSN ejecta, but it only carries a small $^{22}$Na abundance of a few parts per million by mass. \cite{amari:09} tentatively proposed that the $^{22}$Na to explain the Ne-E(L) component was produced in the O/Ne zone, but a quantitative solution is still problematic. 
This is because it is extremely unlikely to form large C-rich dust in O-rich environments \citep[e.g.,][]{ebel:01} and C-rich grains with typical CCSN signatures are not generally consistent with abundances from O-rich material \citep[][]{lin:10}. More recently, \cite{sieverding:18} reconsidered the additional production of $^{22}$Na by neutrino spallation in the more external (and less O-rich) O/C zone, mostly via the $^{22}$Ne($\nu$$_{\rm e}$, e$^-$)$^{22}$Na reaction, locally increasing the $^{22}$Na for some of the models up to a few 10$^{-6}$ in mass fraction. While such quantities would be compatible with the typical $^{22}$Na abundances in the O/Ne zone and they would not significantly change the total $^{22}$Na yields ejected by the CCSN explosion, \cite{sieverding:18} concluded that the much smaller O/C ratio found in the O/C zone compared to deeper regions could allow to form C-rich mixtures between the O/C zone and the C-rich and $^{22}$Na-poor external He/C zone, without overly diluting $^{22}$Na. In any case, all these models predict local peak $^{22}$Na abundances of up to a few 10$^{-6}$ in mass fraction. 

\cite{pignatari:15} presented new CCSN models that experienced late H
ingestion into the He shell shortly before the CCSN explosion, with the CCSN shock still finding traces of H in these He-rich and C-rich regions. The resulting nucleosynthesis is characterized by strong excesses in $^{15}$N and $^{26}$Al, with C isotopic ratios varying by orders of magnitudes in different C-rich local mixtures of CCSN material. 
Achievements and challenges in reproducing presolar grain signatures are highlighted by several works where this first generation of CCSN models affected by H ingestion are adopted \citep[e.g., ][]{liu:16, liu:17, liu:18, hoppe:19, schofield:22, hoppe:23, liu:24, hoppe:24}.
\cite{pignatari:15} also identified a local production of $^{22}$Na in the He-shell layers of these models up to 10$\%$ in mass fraction during the explosion, exceeding by orders of magnitudes any other known nucleosynthesis pathway to make this isotope in CCSNe. Besides mentioning the potential relevance for graphites, however, the impact of such a production on CCSN light curves and observations has not been fully explored. 

A well-known signature of $^{22}$Na decay is the $\gamma$-ray emission at 1274.53 keV \citep[e.g.,][]{diehl:98}. \cite{iyudin:10} reported 
a possible COMPTEL
detection at 4-$\sigma$ confidence level of the $^{22}$Na $\gamma$-ray line from the Nova Cassiopeia 1995, with an estimated $^{22}$Na abundance in the order of 10$^{-7}$ M$_{\odot}$. 
Such 
an observation 
would be consistent with stellar theoretical simulations of novae from typical ONe white dwarfs, where the nova events are activated from the H burning thermal runaway on top of the white dwarf accreting material from a closeby stellar companion \citep[e.g., ][]{jose:98, denissenkov:14, starrfield:16,  leung:22, diehl:22}. However, differently from typical rapid and more massive ONeMg novae, Nova Cassiopeia 1995 had very slow-expanding ejecta and low progenitor mass (M$\sim$0.6 M$_{\odot}$), making its origin still unclear \citep[][]{takeda:18}. V1405 Cas is another more recent example of slow and low-mass Ne nova \citep[][]{munari:22, taguchi:23}.
\cite{shara:94} and \cite{shara:94a} predicted that white dwarfs smaller than typical ONeMg Novae with varying mass accretion stages could have generated slow-ejecting novae (like e.g., Nova Cassiopea 1995), providing ideal candidates for $^{22}$Na detection. However, limited simulations are available exploring these regimes \citep[e.g.,][]{gilpons:03}. For other novae in the literature, only upper limits are available for $\gamma$-ray emission from $^{22}$Na decay \citep[e.g.,][]{siegert:21}.

The detection of the $^{22}$Na $\gamma$-ray emission was never reported (or expected) from CCSNe. In fact, for typical CCSN models without H ingestion, the predicted emission is orders or magnitude lower than COMPTEL or INTEGRAL detection limits \citep[][and references therein]{sieverding:18}. 

The decay of radioactive nuclei produced by the explosion are fundamental sources powering the late-time bolometric light curve of CCSNe. \cite{woosley:89} studied the evolution of the SN1987A light curve considering the contribution of $^{56}$Co ($T$$_{1/2}$ = 77 days), $^{57}$Co ($T$$_{1/2}$ = 272 days), $^{44}$Ti ($T$$_{1/2}$ = 60 years) and of $^{22}$Na. However, with a typical CCSN $^{22}$Na yield of a few 10$^{-6}$ M$_{\odot}$, its contribution to the light curve was found to be marginal or negligible at any time. \cite{timmes:96} found a similar result for SN1987A, where the contribution from the decay of $^{60}$Co ($T$$_{1/2}$ = 5.3 years) was also taken into account in addition to the radioactive sources mentioned above. Therefore, more recent works do not even consider $^{22}$Na as a relevant source powering the CCSN light curve \citep[e.g.,][]{seitenzahl:14}, and the analysis by \cite{sieverding:18} agrees with such an approach. In this work we will show that CCSN models carrying the $^{22}$Na-rich signature of H ingestion challenge those conclusions, and $^{22}$Na may also be relevant for the late-time CCSN light curve.

\subsection{How much $^{22}$Na is required to make the Ne-E(L) component?}
\label{sec:na22_estimate}

An open question is what would be the $^{22}$Na amount required in the parent CCSN ejecta to explain the existence of the Ne-E(L) component in 
graphites, and this is indeed a difficult issue for which we can only derive some general indications.
The most $^{22}$Ne-rich graphites known, with a presumed CCSN origin, are the Murchison grain KFB1-161 with about 1.7$\times$10$^{-2}$ cm$^3$ STP/g\footnote{Units for volume in gaseous form per gram at standard temperature and pressure.} \citep[][]{nichols:92} and the Orgueil grain OR1d-8m-7 with 1.9$\times$10$^{-2}$ cm$^3$ STP/g \citep[][]{heck:18}. Therefore, we could consider a maximum Ne-E(L) concentration of 2.0$\times$10$^{-2}$ cm$^3$ STP/g as a reference. For graphites, this concentration would correspond to an atomic $^{22}$Na/$^{12}$C ratio of 
1.1$\times$10$^{-5}$, where we considered that the molar volume of an ideal gas is 22400 cm$^3$ STP/mol, 
and that there are 5.02$\times$10$^{22}$ atoms in 1g of $^{12}$C. The $^{22}$Na condensation time in dust (given that $^{22}$Na is short-lived) and the fractionation factor C/Na during condensation should also then be taken into account, and both quantities are difficult to estimate. 

Regarding the condensation time, for SiC-X grains \cite{ott:19} estimated a timescale of 20 yr, which is 7-8 times the half-life of $^{22}$Na. If we assume that graphites would have a comparable condensation time in CCSN ejecta, the $^{22}$Na abundance in the C-rich mixture where the grains condensed from would have been at least 128 times higher than that measured in grains KFB1-161 and OR1d-8m-7 at the time of the CCSN explosion, i.e., $^{22}$Na/$^{12}$C $\sim$ 0.001. 
There are no estimates available to date for the C/Na fractionation factor in graphites, but we may expect it to be at least much larger than unity since graphite condenses at high temperatures 
\citep[e.g.,][]{lodders:95}. 
Indeed, graphite grains tend to condense at about 1600-2000 K, where the exact temperatures depend on pressure and gas compositions \citep[e.g.,][]{bernatowicz:96}. 
Na in particular is a volatile element, and it would start to condense at much lower temperatures, while graphite grains already started to form. Thus, we could reasonably expect that not all Na in the gas would condense into graphite grains, but that the $^{22}$Na effectively condensed in graphite grains would be instead much lower than that in the gas of the original composition. At the moment, it is unknown how much Na in the original gas would condense into graphite grains. 
To give an idea about how these factors would change significantly for different elements, for presolar SiC condensation we derived an N/C fractionation of a factor of 50 \citep[][]{hoppe:18} and an S/Si fractionation of a factor of 10$^4$ \citep[][]{pignatari:13a}.

Based on these considerations, we may use $^{22}$Na/$^{12}$C $\sim$ 0.001 as a lower limit for the initial CCSN material mixture from which KFB1-161 and 
OR1d-8m-7 condensed (i.e., using the unrealistically low C/Na fractionation of 1). In the C-rich He/C zone ejecta, the $^{12}$C abundance typically varies between a few up to 20-30 per cent in the deepest parts of the former He shell \citep[e.g.,][]{denhartogh:22} and in the C/Si zone \citep[][]{pignatari:13}. Therefore, using a $^{12}$C abundance of 10 per cent as representative of the C-rich CCSN mixture from where C-rich grains would condense, a $^{22}$Na/$^{12}$C $\sim$ 0.001 would correspond to a $^{22}$Na average abundance of $\sim$ 0.0001. 
This concentration is at least two orders of magnitude higher than the typical $^{22}$Na abundances of a few parts per million predicted in O-rich CCSN ejecta 
in typical CCSN ejecta, even without considering a necessary dilution with the $^{22}$Na-poor C-rich material of the He/C zone. So, 
we may conclude that the $^{22}$Na production in the O/Ne region or in the O/C zone cannot be responsible for the Ne-E(L) component. 
In this work we will instead show that the amount of $^{22}$Na made in the He shell of CCSN models with H ingestion is compatible with the estimates presented here.  \\

This work is organized as follows. Section \ref{sec:models} describes the stellar models and the code used to obtain the supernova light curves. 
The results are discussed in Sections \ref{sec:light_curves}, \ref{sec:gamma} and \ref{sec: positron}, and the conclusions are given in Section \ref{sec:conclusions}.

\section{The production of $^{22}$Na and $^{26}$Al in stellar models with late H ingestion} 
\label{sec:models}

Twelve CCSN models with H ingestion were calculated from a progenitor with initial mass M = 25M$_{\odot}$ and initial metallicity Z = 0.02\footnote{The reference solar elemental abundance adopted was \cite{grevesse:93}, but using a more recent solar isotopic distribution by \cite{lodders:03}.}. 
The stellar progenitor was calculated using the GENEC code \citep[][]{eggenberger:08}. Rotation and magnetic fields were not included \citep[][]{pignatari:16}. 
The six most energetic models are 25T-H, 25T-H5, 25T-H10, 25T-H20, 25T-H50, and 25T-H500, already introduced by \cite{pignatari:15}. The 25T-H model carries the H enrichment in the whole He-rich shell material, obtained in the 25 M$_{\odot}$ progenitor \citep[about 1.2\% of H][]{pignatari:15,pignatari:16}, 
while in the other models the H abundance in the He-shell is reduced by a factor of 5, 10, 20, 50, and 500. The CCSN explosion of the 25T models is produced by artificially increasing the temperature and density peak of the original 25M$_{\odot}$ model, to reproduce the explosive conditions of a 15 M$_{\odot}$ star in the He shell layers \citep[][]{pignatari:15, schofield:22}. The other six models considered here (25av-H, 25av-H5, 25av-H10, 25av-H20, 25av-H50, and 25av-H500) were designed to have peak temperature and density peaks at the midpoint between the 25T models and the original CCSN 25 M$_{\odot}$ model \citep[][]{schofield:22}.

From comparing the amount of $^{26}$Al produced and the $^{26}$Al/$^{27}$Al ratio found in these models with C-rich presolar grains from CCSNe, \cite{schofield:22} showed that the 25T-H, 25T-H5, 25T-H10 and 25av-H, 25av-H5, 25av-H10 models produce amounts of $^{26}$Al compatible with observations, up to the largest $^{26}$Al/$^{27}$Al ratio measured being larger than one \citep[][]{liu:24, hoppe:24}. 

\begin{figure*}[t]
            \centering
            \includegraphics[width=0.49\linewidth]{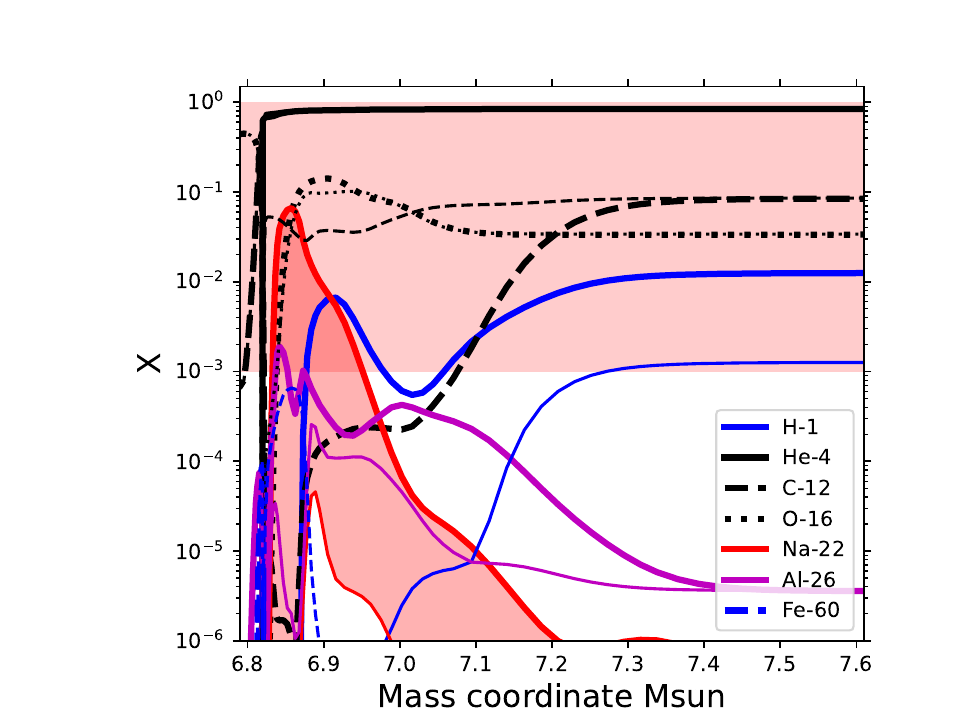}
            \includegraphics[width=0.49\linewidth]{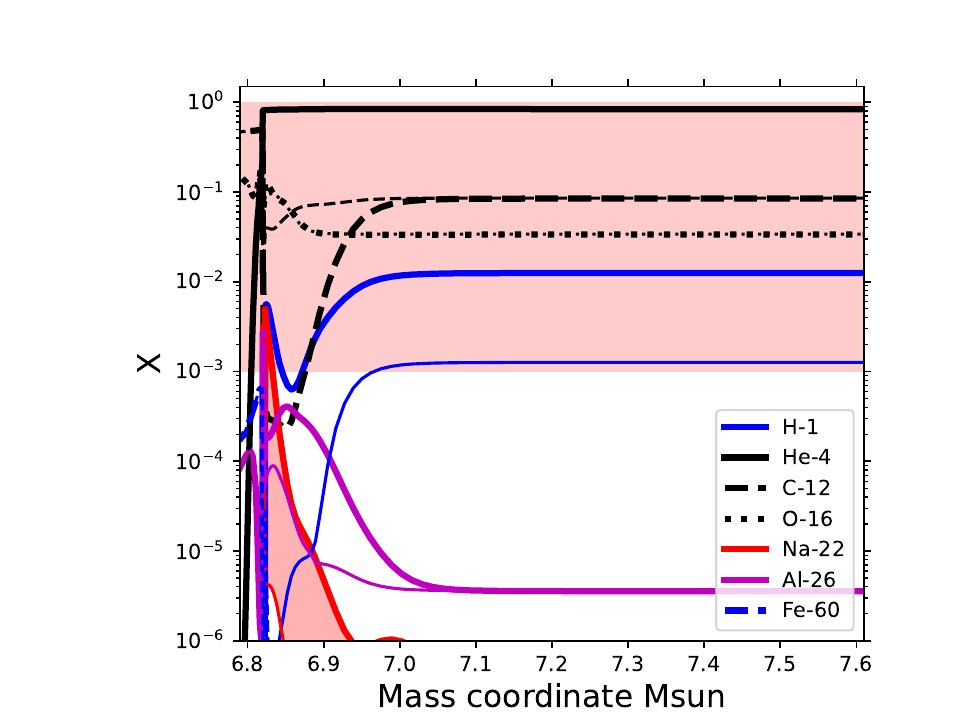}\par
            \includegraphics[width=0.49\linewidth]{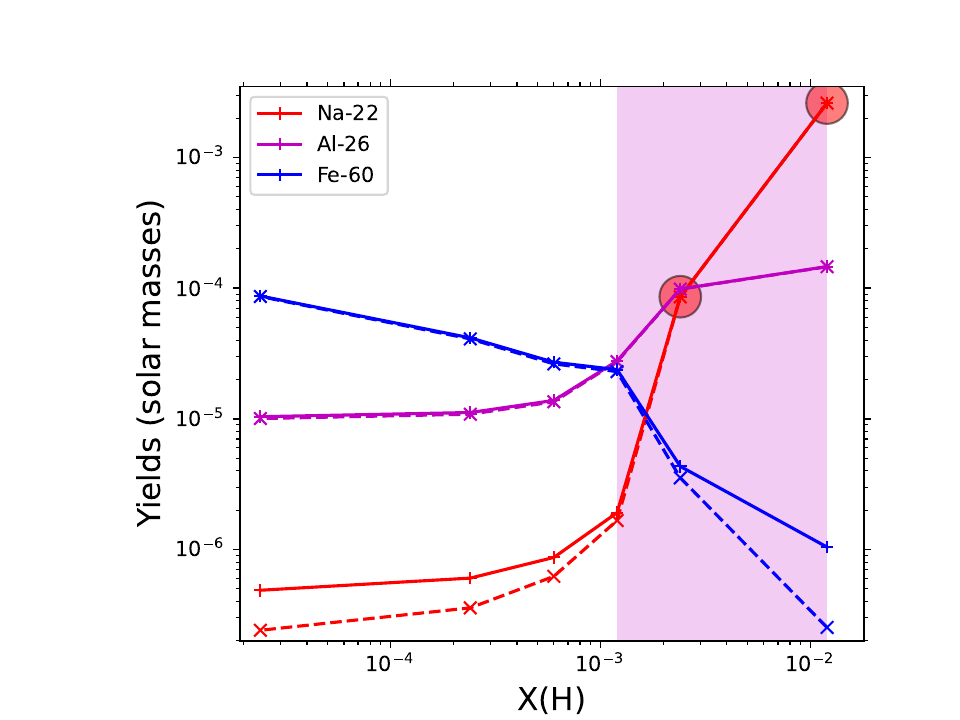}
            \includegraphics[width=0.49\linewidth]{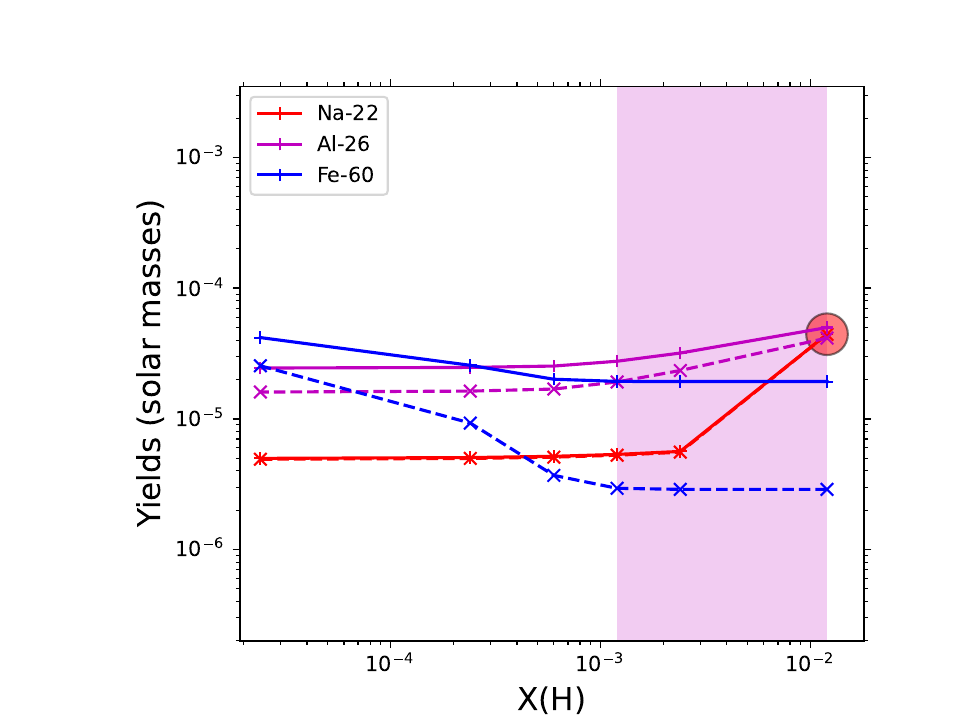}\par
            \caption{Top left panel: isotopic abundances (in mass fraction) in the He-shell ejecta of the CCSN models 25T-H and 25T-H10 (thick and thin lines, respectively). 
            The horizontal red colored band highlights the $^{22}$Na production range compatible with the Ne-E(L) component (see text for details); Top right panel: As the top left panel, but for the 25av-H and 25av-H10 (thick and thin lines, respectively). 
            Bottom left panel: Total $^{22}$Na yields in solar masses (continuous lines) and from the He-shell ejecta only (dashed lines) for all the 25T set models with respect to the available amount of hydrogen at the onset of the CCSN explosion in the He shell. For comparison, the same data are provided for $^{26}$Al and $^{60}$Fe.  The yields of the models consistent with the Ne-E(L) component are highlighted with large red circles. The vertical magenta colored area highlights the production variation between the 25T-H and the 25T-H10 models. In this area, the total $^{22}$Na yields are always dominated by the He-shell ejecta. Bottom right panel: As the bottom left panel, but for the 25av set of CCSN models.}
            \label{fig:na22_yields}
\end{figure*}

The post-explosive abundance profiles of $^{22}$Na, $^{26}$Al and $^{60}$Fe from our selected models in the He shell material are shown in the upper panels of Figure \ref{fig:na22_yields}. 
As discussed by \cite{pignatari:15}, the presence of protons in the He shell at the passage of the CCSN shock tends to suppress the $^{22}$Ne($\alpha$,n)$^{25}$Mg neutron source reaction and the following neutron burst, since the $^{22}$Ne(p,$\gamma$)$^{23}$Na reaction has a much higher efficiency. 
Therefore, $^{60}$Fe and other neutron-capture products are not produced in these conditions. On the contrary, the activation of proton captures in the explosive He shell allows to make significant amounts of both $^{22}$Na and $^{26}$Al. In particular, in Figure \ref{fig:na22_yields} the $^{22}$Na abundance peak is between almost 0.1 (25T-H) and about 10$^{-5}$ (25av-H10), in mass fraction, in these CCSN ejecta (the area between the two $^{22}$Na abundance curves is highlighted in red). The corresponding total yields ejected are provided in the lower panels of the same figure. 
For the models of set 25T that are consistent with the highest $^{26}$Al measurements in presolar grains, i.e., 25T-H, 25T-H5, 25T-H10 \citep[][]{schofield:22}, the total $^{60}$Fe yields decrease by up to two orders of magnitude compared to the model with the lowest amount of H available (25T-H500), $^{26}$Al increases by up to an order of magnitude and $^{22}$Na by more than three orders of magnitudes (left panel). For the 25av models (right panel), the most varying isotope is $^{22}$Na, where the yield obtained with 
the 25av-H model is an order of magnitude larger than 25av-H5 and 25av-H10 (see the red-shaded area between the two $^{22}$Na curves in the figure).    
  
Figure \ref{fig:na22_yields} shows that the $^{22}$Na yields are of the order of a few 10$^{-6}$ M$_{\odot}$ for 25T-H10, 25av-H10 and 25av-H5, comparable to the typical production from CCSN models without proton ingestion. With respect to previous models \citep[e.g.,][]{rauscher:02, chieffi:04, sieverding:18}, the main difference is that the $^{22}$Na production (with local abundance peaks between a few 10$^{-7}$ and a few 10$^{-6}$) would be directly located in the C-rich He shell material. In any case, as discussed in \S \ref{sec:na22_estimate}, these abundances are by orders of magnitudes too small to explain the Ne-E(L) component. Models 25T-H, 25T-H5 and 25av-H show instead local abundances larger than one per mil in the He-shell material. In the figure, we indicated a $^{22}$Na reference abundance = 0.001 as a qualitative indicator of the minimum local production compatible with the Ne-E(L) component. This would be roughly compatible with the $^{22}$Na $>$ 0.0001 lower limit derived in \S \ref{sec:na22_estimate}, accounting for a factor of 10 dilution between the most $^{22}$Na-rich component and the $^{22}$Na-poor C-rich material in the mixture of CCSN ejecta where graphites condense. Notice that such a dilution factor is an arbitrary choice. At this stage we cannot yet derive a clear recipe for this, because of the uncertainties related to the $^{22}$Na condensation into grains (see the discussion is section \ref{sec:na22_estimate}). Indeed, because of elemental fractionation all $^{22}$Na would likely not condense into graphite grains with the original relative abundances. 
Furthermore, we could realistically assume a wide grain-to-grain variations if we consider the isotopic anomalies for indicative elements like Si and Ca \citep[e.g.,][]{besmehn:03, pignatari:13, schofield:22}. For instance, dilution factors down to a factor of 200 were considered by \cite{xu:15} for the different components ejected, to generate mixtures of CCSN material to reproduce presolar grain abundances. 

A factor of 10 dilution was assumed here to take all these considerations into account. 

In Figure \ref{fig:na22_yields}, the extra production of $^{22}$Na due to the overlapping contribution of proton ingestion and the CCSN explosion results in $^{22}$Na yields between about 4$\times$10$^{-5}$ and 3$\times$10$^{-3}$ M$_{\odot}$ (Figure \ref{fig:na22_yields}), between one and three orders of magnitude larger than the typical CCSN production dominated by the O/Ne zone and/or by the O/C zone contributions \cite[e.g.,][]{sieverding:18}.

\begin{figure*}[t]
            \centering
            \includegraphics[width=0.49\linewidth]{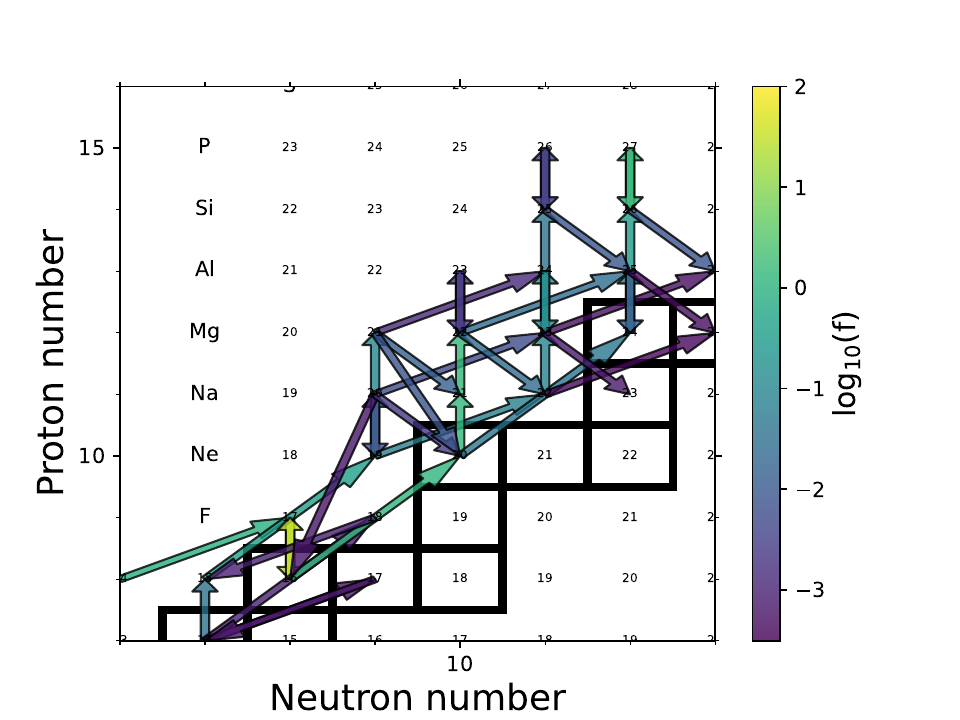}
            \includegraphics[width=0.49\linewidth]{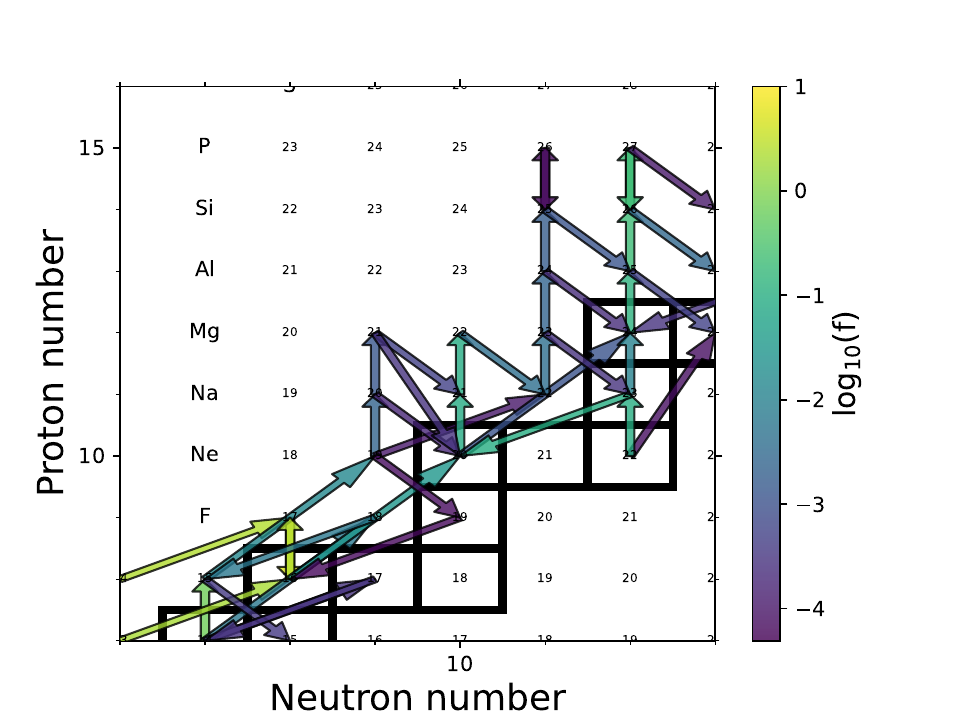}\par
            \includegraphics[width=0.49\linewidth]{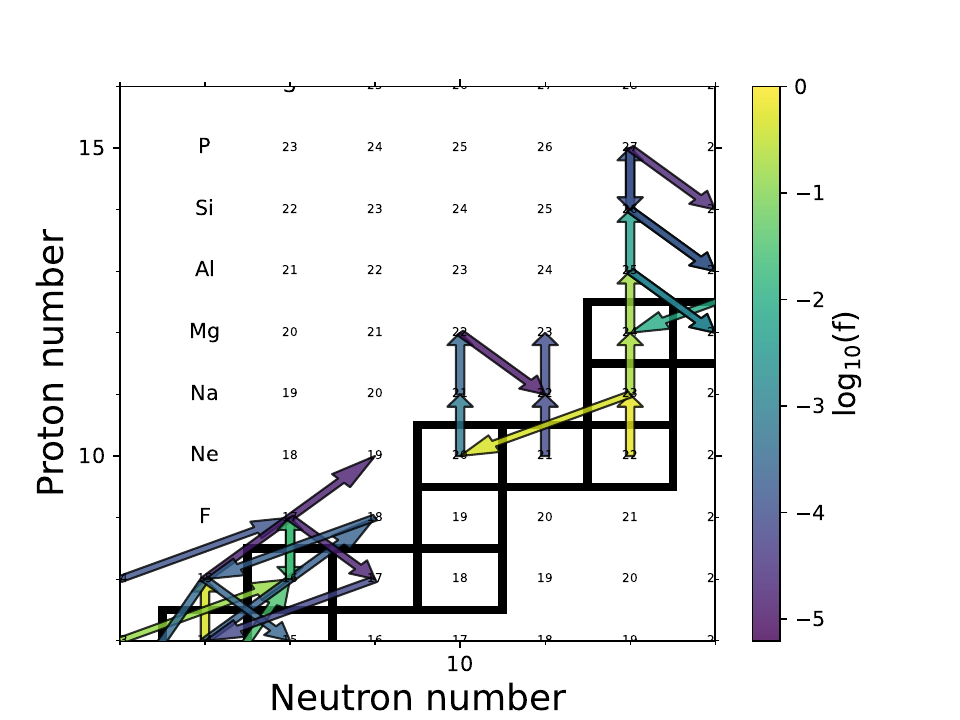}
            \includegraphics[width=0.49\linewidth]{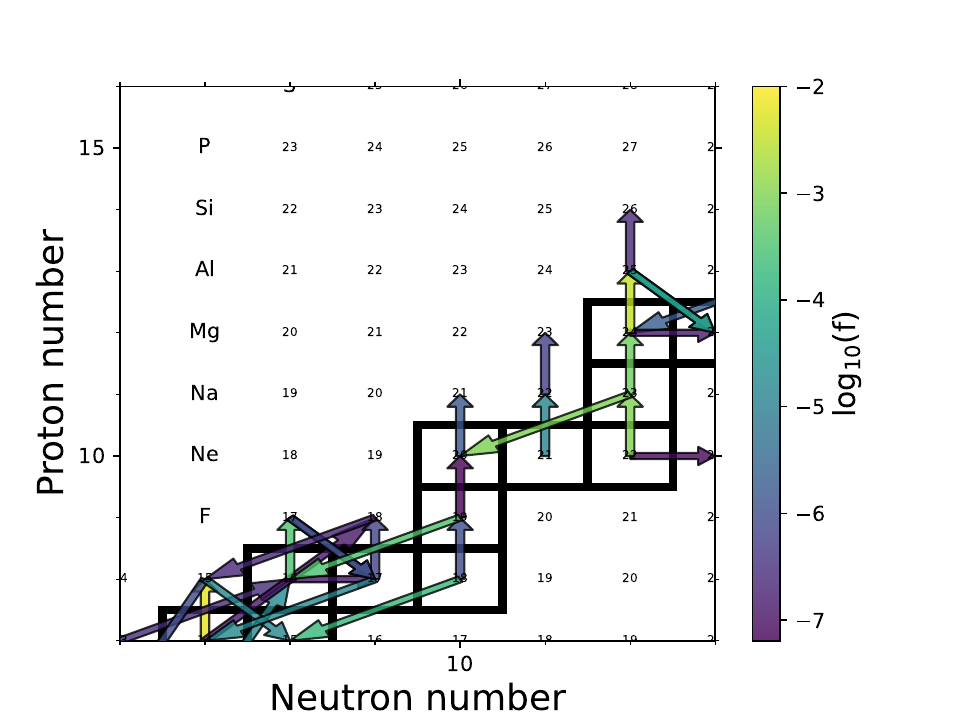}\par
            \caption{Integrated nucleosynthesis fluxes (log$_{10}([\delta$Y$_{\rm i}$/$\delta$t]$_{\rm j})$, showing the cumulative variation of the abundance Y$_{\rm i}$ = X$_{\rm i}$/A$_{\rm i}$ due to the reaction j) at 47 seconds after the peak explosion temperature at mass coordinates 6.855 M$_{\odot}$ (top left, CCSN shock peak temperature T$_{\rm peak}$ = 1.1 GK), 6.908 M$_{\odot}$ (top right, T$_{\rm peak}$ = 0.8 GK), 7.038 M$_{\odot}$ (bottom left, T$_{\rm peak}$ = 0.55 GK), 7.333 M$_{\odot}$ (bottom right, T$_{\rm peak}$ = 0.33 GK) of the model 25T-H, in the regions of interest for the $^{22}$Na production. These mass coordinates correspond to different production efficiency of $^{22}$Na (see Figure \ref{fig:na22_yields}, upper left panel). The arrow 
            color scale on the right correspond to the flux strength. 
            Heavy-lined boxes correspond to the stable isotopes.}
            \label{fig:na22_flux}
    \end{figure*}

The dominating nucleosynthesis fluxes of the 25T-H CCSN ejecta are shown in Figure \ref{fig:na22_flux} for different mass coordinates. In the He shell region where a large range of $^{22}$Na values are obtained (see Figure \ref{fig:na22_yields}, thick lines, top left panel), the $^{22}$Na peak abundance in mass of 6.5 $\times$ 10$^{-2}$ is shown at mass coordinate 6.855 M$_{\odot}$ (Figure \ref{fig:na22_flux}, top left), 
7 $\times$ 10$^{-3}$ at 6.908 M$_{\odot}$ (top right), 
2.7 $\times$ 10$^{-5}$ at 7.038 M$_{\odot}$ (bottom left) 
and 1 $\times$ 10$^{-6}$ at 7.333 M$_{\odot}$ (bottom right). 
The temperature peaks reached in these regions vary between 1.1 GK and 0.3 GK. 

The $^{22}$Na nucleosynthesis in these conditions varies greatly from one mass location to the other, as shown in Figure \ref{fig:na22_flux}. For instance, in the hottest conditions (upper left) a number of ($\alpha$,p) reactions are activated together with proton captures, similarly to X-ray burst conditions in accreting neutron stars \citep[e.g.,][]{wiescher:99}. In this specific trajectory, the $^{19}$Ne($\alpha$,p)$^{22}$Na flux is clearly relevant for the production of $^{22}$Na, and the rate currently used is based on theory. However, in a simple test reducing its rate by a factor of a hundred, there was only a 
3\% variation on the final $^{22}$Na abundance calculated. In this case the alternative $^{19}$Ne(p,$\gamma$)$^{20}$Na(p,$\gamma$)$^{21}$Mg flow depletes $^{19}$Ne, feeding the production path $^{20}$Ne(p,$\gamma$)$^{21}$Na(p,$\gamma$)$^{22}$Mg via the $^{20}$Na decay to $^{20}$Ne.
$^{22}$Mg decays to $^{22}$Na, although the $^{22}$Mg($\alpha$,p)$^{25}$Al reaction also destroys $^{22}$Mg. While we expect the rate of this reaction to affect the $^{22}$Na production peak reached in our CCSN models, its experimental value is still a matter of debate.  
The $^{22}$Mg($\alpha$,p)$^{25}$Al reaction has been studied recently in the context of X-ray bursts and rp-process nucleosynthesis. In particular, three measurements were performed using radioactive ion beams and complementary techniques. \cite{randhawa:20} 
performed the first direct measurement using an active-target time-projection chamber (AT-TPC) system. \cite{hu:21} 
used $^{25}$Al+p elastic and inelastic scattering to extract resonance information of states in $^{26}$Si with the thick-target yield in inverse kinematics. \cite{jayatissa:23} employed the MUSIC active-target system to directly measure the reaction, covering a similar energy range as \cite{randhawa:20}. Surprisingly, the three results are in disagreement with each other, beyond the given errors, and with the theoretical predictions, based on the statistical nuclear reaction mechanism. 

Uncertainties of other nuclear rates are also relevant for $^{22}$Na nucleosynthesis in stars, 
for example the $^{21}$Na(p,$\gamma$)$^{22}$Mg and the $^{22}$Na(p,$\gamma$)$^{23}$Mg reactions. 
For many conditions, the $^{21}$Na(p,$\gamma$)$^{22}$Mg production pathway flow is activated. However, this reaction has been studied extensively in the context of novae and X-ray bursts. In particular, direct measurements with a $^{21}$Na beam using the DRAGON recoil separator \citep[][]{dauria:04} 
covered the temperature range of relevance here and the reaction rate is considered well-constrained.
The definition of the $^{22}$Na(p,$\gamma$)$^{23}$Mg reaction rate is instead still controversial. A recent measurement of the lifetime of the relevant 7785 keV state in $^{23}$Mg by \cite{fougeres:23}, resulted in a resonance strength compatible with the upper limit set by a previous direct measurement \citep[][]{seuthe:90}, but disagrees with two other direct measurements \citep[][]{sallaska:10, stegmuller:96}. 
Therefore, the $^{22}$Na(p,$\gamma$)$^{23}$Mg reaction is still uncertain in the relevant temperature regions for Novae (0.1-0.4 GK), and for both X-ray bursts and this study (0.6-1.1 GK).
New measurements for this reaction are needed to elucidate the disagreements. 

A detailed study about the impact of nuclear reaction rates and their uncertainties on the $^{22}$Na production in CCSN models from massive stars affected by H ingestion, is needed in the future. 

\section{Late CCSN light curves: $^{22}$Na or $^{44}$Ti?}
\label{sec:light_curves}


There is growing evidence that many supernovae have late-time power sources starting from a few hundreds days after the CCSN explosion, the brightest of which require shock interactions or magnetars~\citep{2024arXiv240504259D,2024arXiv240813844D}. But radioactive decay can also contribute to the late time emission.

Even though the decay half-life of $^{22}$Na is ideally suited to powering light-curves at the 1-2\,years timescale, it is currently not considered as a relevant source for CCSN light curves. However, the yields shown in Figure \ref{fig:na22_yields} may significantly change the picture. In particular, the models compatible with the Ne-E(L) component in presolar grains (25T-H, 25T-H5 and 25av-H) have $^{22}$Na yields of 2.61$\times$10$^{-3}$ M$_{\odot}$, 8.60$\times$10$^{-5}$ M$_{\odot}$ and 4.44$\times$10$^{-5}$ M$_{\odot}$ respectively. These are between twenty and a thousand times larger than the values used by e.g., \cite{woosley:89} to calculate the SN1987A light curve.

To study the impact of these values on the bolometric light curve of a CCSN like SN1987A, we have used a python version of the open-source code \verb|snlite|\footnote{\url{https://cococubed.com/code_pages/snlite.shtml}}. We used the $^{56}$Co, $^{57}$Co, $^{44}$Ti and $^{60}$Co abundances by \cite{timmes:96} of 0.069 M$_{\odot}$, 0.0033 M$_{\odot}$, 10$^{-4}$ M$_{\odot}$, 2$\times$10$^{-5}$ M$_{\odot}$, respectively. Their production in the He shell is negligible compared to more internal CCSN ejecta, and their yields are not 
affected by H ingestion. At these late times after the explosion,  additional radioactive isotopes may also contribute to the total light curve and their contribution should be studied in more detail \citep[e.g., $^{56,59}$Ni, $^{48,49}$V and $^{51}$Cr,][]{timmes:19}. 

Since we are only interested here to explore the relative contribution of the different radioactive isotopes and we are not directly comparing with observations, we will consider the UVOIR pseudo-bolometric light curve, i.e., we assume that the emitting material where the radioactive isotopes are located is still optically thick so that the $\gamma$-rays are thermalized. Therefore, we are not applying any reduction factor to the total light-curve \citep[e.g.,][]{seitenzahl:14}, and we are not taking into account that $\gamma$-rays will become fully untrapped between 100 and 1000 days after the explosion while positrons can still be trapped depending on the magnetic field distribution in the CCSN remnant and help power the light-curve \citep[see e.g.,][and discussion in Section~\ref{sec: positron}]{kozma:98}. 
Notice that a direct comparison between the observed bolometric luminosities between 500 and 1500 days of SN1987A \citep[][]{suntzeff:92} and the theoretical bolometric light curve was already made by \cite{timmes:96}, showing that such an approximation still \textbf{allows} to reproduce quite well observations in the give time range after the CCSN explosion. 

In this section we mostly focus instead in comparing the relative contribution between $^{22}$Na and other relevant radioactive species, deferring a detailed comparison with observations to future papers using more advanced light curve modeling. However, the decline of the late-time light curve is expected to be dominated by the half-life of the radioactive decay.  If this late-time light curve is dominated by $^{22}$Na, we expect a much shallower decay than that produced by the 77 days half-life of $^{56}$Co.  This shallow light-curve may explain the late-time evolution of some supernovae, e.g. SN 2020jvf \citep[][]{sollerman:21}. If this scenario would be confirmed, than the light-curve decline could be used to compare the relative contributions by $^{56}$Ni and $^{22}$Na.

The values of the effective opacities are $\kappa_{56}$ = 0.033 cm$^{2}$ g$^{-1}$, $\kappa_{57}$ = 2.4 $\times$ $\kappa_{56}$ and $\kappa_{44}$ = $\kappa_{22}$ = $\kappa_{60}$ = 0.04 cm$^{2}$ g$^{-1}$, where the indices of the effective opacities refer to the mass number of the respective radioisotopes \citep[][]{woosley:89, timmes:96}. 

\begin{figure}[ht!]
\centering
\includegraphics[width=\columnwidth]{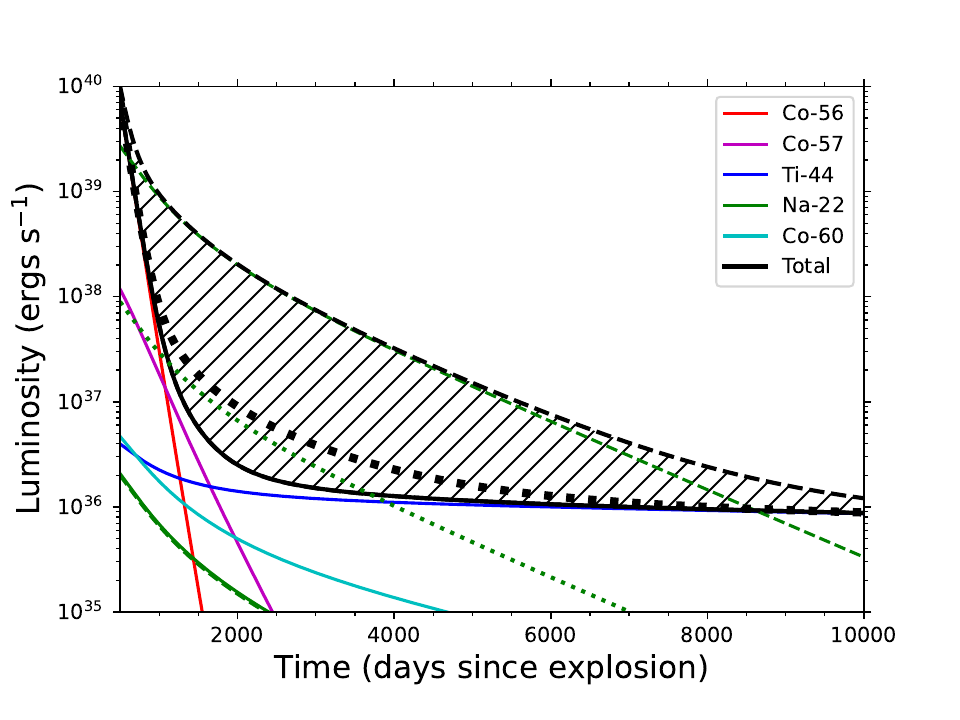}
\includegraphics[width=\columnwidth]{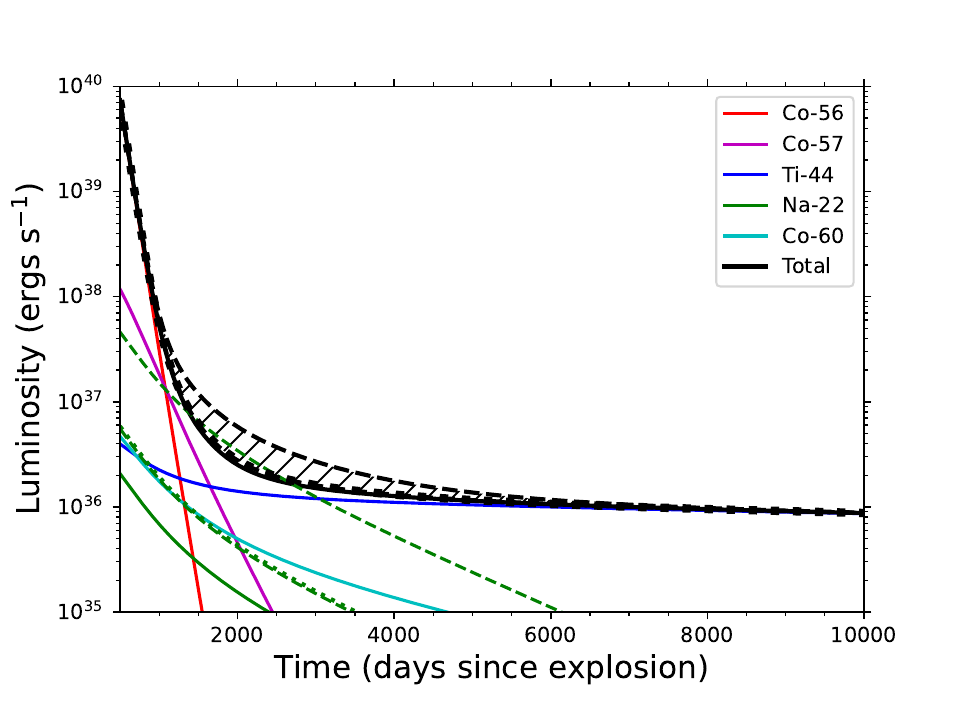}
\caption{Top panel: The UVOIR pseudo-bolometric light curve 
is shown between 500 and 5000 days after the explosion \citep[black continuous line,][]{timmes:96}. The relative contribution from  the decay of $^{56}$Co, $^{57}$Co, $^{44}$Ti, $^{60}$Co and $^{22}$Na are considered using previous default abundances. In particular, the partial $^{22}$Na curve obtained using the default value of 2$\times$10$^{-6}$ M$_{\odot}$ from \cite{timmes:96} is also shown (green continuous line). For the range of $^{22}$Na production obtained in the 25T-H (highest green dashed line), 25T-H5 (green dotted line) and 25T-H10 models (lowest green dashed line),  
the corresponding total pseudo-bolometric light curves are shown as a higher black dashed line, black dotted line and lower black dashed line respectively.  
Their variation due to the $^{22}$Na production range is highlighted as a black striped area. Bottom panel: Same as the top panel, but considering the $^{22}$Na production of the 25av-H, 25av-H5 and 25av-H10 models.
}
\label{fig:na22_lightcurves}
\end{figure}

The results are shown in Figure \ref{fig:na22_lightcurves} for models 25T-H, 25T-H5 and 25T-H10 (top panel) and 25av-H, 25av-H5 and 25av-H10 (bottom panel). 
The models not compatible with the Ne-E(L) component do not show a significant impact on the CCSN light curves. Instead, for the most $^{22}$Na-rich models compatible with the Ne-E(L) component (25T-H, 25T-H5 and 25av-H), the total light curve is affected by the $^{22}$Na decay contributing during both the radioactive tail of $^{57}$Co and later the radioactive tail of $^{44}$Ti \citep[e.g.,][]{leibundgut:03}. According to these models, the $^{22}$Na contribution will fade away only after thousands of days, when eventually only $^{44}$Ti remains to contribute, due to its longer half-life. 
With a $^{44}$Ti abundance of 10$^{-4}$ M$_{\odot}$, this will happen after about 5000 days (25av-H), 6000 days (25T-H5), and 10000 days (25T-H). 

The production of $^{44}$Ti in CCSNe and its relevance for the SN light curve has been extensively discussed in the literature \citep[e.g.,][]{magkotsios:10, grebenev:12, seitenzahl:14, sieverding:18, subedi:20, hermansen:20}. \cite{the:06} in particular highlighted a potential (and unexpected from models) large variation of $^{44}$Ti yields in different CCSN remnants. Instead, because of the large range of conditions that could be triggered by H-ingestion events in different CCSN progenitors, we expect a variation by orders of magnitudes of $^{22}$Na abundance yields. This would naturally cause a significant variability of the late light curves between different CCSN, which could be due to the different $^{22}$Na amounts, without changing typical $^{44}$Ti CCSN yields. We do not know if CCSN material as for SN1987A and CasA was affected by H-ingestion events. Nevertheless, based on current CCSN models we conclude that $^{22}$Na production compatible with the Ne-E(L) component in presolar graphite grains would affect the radioactive tails of CCSN light curves from about 500 days after the CCSN explosion. 

\section{$\gamma$-ray emission from $^{22}$Na: too old but not too far} 
\label{sec:gamma}


While the $^{22}$Na $\gamma$-ray emission at 1274.53 keV is orders of magnitude lower than the COMPTEL and INTEGRAL detection limits if typical CCSN $^{22}$Na yield of a few 10$^{-6}$ M$_{\odot}$ are used, this may not be the case for the H-ingestion models.  
Therefore, we present the predictions for the $^{22}$Na emission line assuming for simplicity that 20\% of the $^{22}$Na $\gamma$-rays are not thermalized and are emitted instead from the system, 
starting from a year after the explosion \citep[][]{sieverding:18}. 

\begin{figure}[ht!]
\centering
\includegraphics[width=\columnwidth]{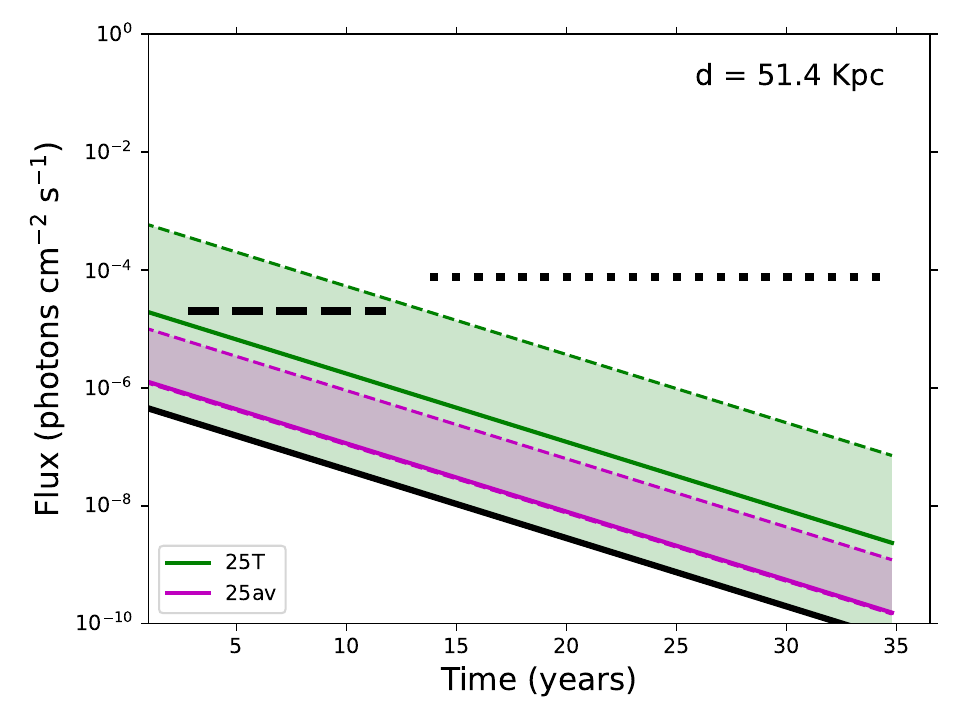}
\includegraphics[width=\columnwidth]{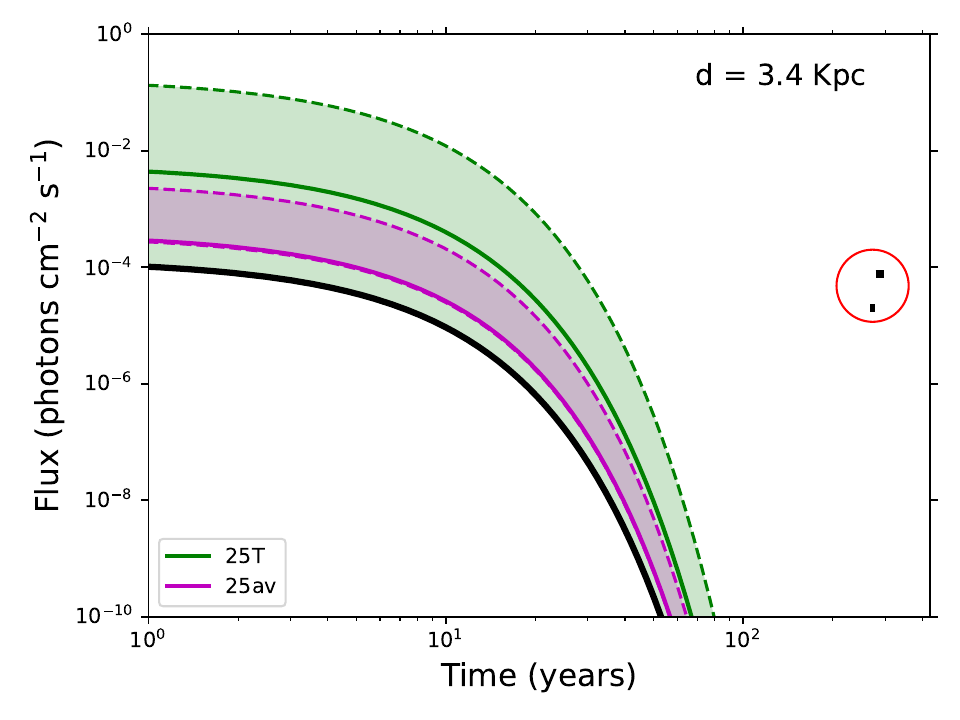}
\caption{Upper Panel: The predicted 1274.53 keV $\gamma$-ray flux due to $^{22}$Na decay is calculated for the SN1987A remnant from the first possible detection of the CCSN explosion until the present day, if we consider the yields from the models 25T-H (higher green dashed line), 25T-H5 (green continous line) and 25T-H10 (lower green dashed line) and the models 25av-H (higher magenta dashed line), 25av-H5 (magenta continous line) and 25av-H10 (lower magenta dashed line). The area between the two sets of curves are colored accordingly. 
A distance of 51.4 Kpc is adopted \citep[][]{panagia:99}. The flux obtained using the default $^{22}$Na abundance from \cite{timmes:96} is reported as a reference (black continuous line). Note that the curves for 25T-H10 and 25av-H10 are not visible, hidden behind the default (black) curve (25T-H10) and the 25av-H curve (25av-H10).
Horizontal lines are the detection limit for COMPTEL \citep[2 $\cdot$ 10$^{-5}$ photons cm$^{-2}$ s$^{-1}$, dashed line][]{iyudin:10} and INTEGRAL \citep[7.6 $\cdot$ 10$^{-5}$ photons cm$^{-2}$ s$^{-1}$, dotted line][]{siegert:18} during their operation time. Lower Panel: The same as in the Upper Panel, but for Cas A, at 3.4 Kpc. 
In this case, COMPTEL and INTEGRAL detection limits appear just like data points (within the red circle) along the longer timeline. 
}
\label{fig:na22_gammaray}
\end{figure}

Figure \ref{fig:na22_gammaray} shows the expected 1274.53 keV $\gamma$-ray flux due to the decay of $^{22}$Na from models 25T and 25av, for a CCSN exploded at the time and distance of SN1987A (top panel) or CasA (bottom panel). For a CCSN at a distance of the order of 50 Kpc, the $^{22}$Na $\gamma$-signature consistent with the Ne-E(L) component would be detectable for up to 10 years, using the COMPTEL detection limit. The detection period would be up to around 40 years for a CCSN at a distance of around 3 Kpc. Concerning SN1987A, even considering the model with the largest $^{22}$Na yields 25T-H, the $^{22}$Na decay signature would have been below the INTEGRAL detection capabilities. A 25T-H signature could have been detectable by COMPTEL, while 25T-H5 model is just below the detection limit at the beginning of its operation. However, in the early stages ($\sim$ year) it is difficult to distinguish the $^{22}$Na line from the $^{56}$Co line at 1238 keV \citep[][]{leising:90}, where the COMPTEL energy resolution at these energies was about 5\% of FWHM, or about 60 keV \citep[][]{schoenfelder:93}. Based on our models, we would expect that at about five years from the explosion there should be up to three orders of magnitude more $^{22}$Na than $^{56}$Co due to its much longer half-life, and therefore any line in this energy window would definitely come from the $^{22}$Na decay (see e.g., in Figure \ref{fig:na22_lightcurves} for the same effect on the light curve). Unfortunately, we were not able to find any SN1987A spectra from COMPTEL at these later stages in the literature. 

Future detectors for $\gamma$-ray astronomy could improve the probability to detect the $^{22}$Na signature from different stellar sources including CCSNe \citep[][]{timmes:19}. The Compton Spectrometer and Imager (COSI) NASA small explorer, scheduled to launch in 2027, will have $\sim 3 \times 10^{-6} {\rm \, photons \, cm^{-2} \, s^{-1}}$ for the 1274.53 keV line expected from the decay of $^{22}$Na~\citep{2023arXiv230812362T}. The detector for the proposed e-ASTROGAM $\gamma$-ray telescope will have a predicted 3.8 $\times$ 10$^{-6}$ photons cm$^{-2}$ s$^{-1}$ for $^{22}$Na detection limit \citep[][]{deangelis:18}.
Even with these sensitivities, COSI and e-ASTROGAM will be unable to detect $^{22}$Na from any known young supernova remnant. However, with new Galactic supernovae they will be able to easily distinguish the $^{22}$Na signature predicted by the models discussed here, and they should also be able to detect the $^{22}$Na signature for supernovae throughout the local Group. 


\section{Decay Products: Positrons produced and CCSN Remnants}
\label{sec: positron}

With the much larger $^{22}$Na yields from our models (see figure \ref{fig:na22_yields} and discussion in Section \ref{sec:models}), a CCSN could produce $5 \times 10^{50} - 5 \times 10^{52}$ positrons from $^{22}$Na decay. With a CCSN rate of 1-2 per century and assuming all follow such high H-ingestion, these models would yield a production rate of $10^{41}-10^{43}$ positrons per second, close to the annihilation rate in the Milky Way of about $5 \times 10^{43}$ positrons s$^{-1}$, considering both the measurements in the galactic bulge and the disk \citep[e.g.,][]{knodlseder:05, siegert:16, siegert:23}.  However, a large fraction of these positrons may be trapped in the CCSN flow, annihilates, powering the CCSN light curve (see Section ~\ref{sec:light_curves}). Whether or not these positrons are trapped in the flow depends among other things upon the configuration and orientation of the magnetic field  
\citep[e.g.,][]{1980ApJ...237L..81C, 1997ASIC..486..273C}.  

The escape fraction of the positrons produced by the $^{56}$Co decay has been studied in detail through comparisons to late-time light-curves for thermonuclear supernovae~\citep[SNIa,][]{1999ApJS..124..503M, 2001ApJ...559.1019M, churazov:15}. On average, for SNIa only a few percent of the positrons would escape. However, based on observations of late-time light-curves (and the lack of late-time energy deposition) it was shown that even a subset of SNIa with a high fraction of positron escape would be sufficient to explain most of the Galactic positron rate \citep{{siegert:16}}. 

The escape fraction of positrons is proportional to the density of the ejected material when the positrons are produced. CCSN are typically characterized by higher-mass ejecta ($>5$ times higher) and lower velocities ($\sim 30$\% lower) compared to SNIa. The densities of the SNIa ejecta at 77\,days (77.236\,days is the half-life of $^{56}$Co) is roughly the same density of the CCSN ejecta at 2.6\,years (the half-life of $^{22}$Na). Given that the escape fraction is roughly proportional to this ejecta density, we expect the escape fraction of $^{22}$Na-decay positrons in CCSN to be similar to the $^{56}$Co-decay positrons in SNIa supernovae~\citep{1999ApJS..124..503M,2001ApJ...559.1019M}.  Hence, unless the magnetic field structure in CCSN is very different than that of SNIa, the positrons produced from the decay of $^{22}$Na should only contribute by a few percent to the positron fraction in the Milky Way. 
Therefore, even with the large concentrations of $^{22}$Na compatible with the Ne-E(L) component, escape fractions larger than 20-30\% (\textbf{i.e. an} order of magnitude larger than the fractions we just estimated in the text) would be needed in order to have positrons from $^{22}$Na decay to significantly contribute to the galactic positron annihilation rate. Additionally, we would expect that only a subset of CCSN progenitors would be affected by a late H ingestion and generate the enhanced $^{22}$Na abundance signature. This would reduce even further the net CCSN contribution to the galactic budget. In conclusion, the current (marginal) CCSN contribution to the positron annihilation rate of the galactic disk could be more like a lower limit of their real contribution. Based on the models presented in this work and on the discussion in this section, there could indeed be an additional ($\lesssim$1\%) contribution from the $^{22}$Na decay. At the moment it is not possible to estimate the absolute contribution, due to the model uncertainties and the unknown occurrence of the $^{22}$Na-rich production compatible with the Ne-E(L) component in presolar graphite grains.  



\section{Conclusions and future perspectives} 
\label{sec:conclusions}

Presolar graphite grains from CCSNe show the initial presence of radiogenic $^{22}$Na, in the form of $^{22}$Ne, identified as the Ne-E(L) component. In typical CCSN models the only part of the ejecta where $^{22}$Na is produced is the inner O-rich O/Ne zone far from the C-rich He-rich zone where graphite grains are expected to form. Furthermore, the expected $^{22}$Na is only a few part per million in mass in these regions, which is too low to explain the measured Ne-E(L) component. Those two points have been unsolved puzzles until now. 

We have analyzed the production of $^{22}$Na in twelve 1D CCSN models from massive star progenitors affected by H ingestion, of which six are already consistent with the $^{26}$Al signature in presolar grains from CCSNe. Compared to typical CCSN models, the $^{22}$Na abundances in the C-rich He shell can be increased by several orders of magnitudes, and the integrated yields may increase up to three orders of magnitude. We find that from the six $^{26}$Al-rich models (i.e. with highest H-ingestion), three can also match the Ne-E(L) component.  

We further discussed the impact of the predicted $^{22}$Na abundances 
on the CCSN light curve, on the $^{22}$Na decay $\gamma$-ray emission at 1274.53 keV and on the potential CCSN contribution to the positron annihilation rate observed in the galactic disk. 

The main conclusions are the following:
\begin{itemize}
\item The minimum $^{22}$Na abundance in a C-rich mixture to explain the Ne-E(L) component detected in presolar graphites is 0.0001. This condition is satisfied by three H-ingestion CCSN models: 25T-H, 25T-H5 and 25av-H. The total $^{22}$Na yields in these models are 2.61$\times$10$^{-3}$ M$_{\odot}$, 8.60$\times$10$^{-5}$ M$_{\odot}$ and 4.44$\times$10$^{-5}$ M$_{\odot}$, respectively. 
\item These models (25T-H, 25T-H5 and 25av-H) show a relevant contribution of the decay of  $^{22}$Na to the CCSN light curve starting from about 500 days to 1500 days. 
Predicted $^{22}$Na abundances compatible with the Ne-E(L) component in presolar graphites may be more relevant to the light curve than the $^{44}$Ti contribution for several years if H-ingestion plays a significant role. Beyond the simple calculations presented here, future comprehensive CCSN light curve simulations should test the impact of the predicted $^{22}$Na abundances, and compare with available observations from different CCSNe.   
\item The decay $\gamma$-ray line at 1274.53 keV predicted from the same models would be detectable from telescopes like INTEGRAL and COMPTEL for up to 40 years after a CCSN explosion at a few Kpc (e.g., CasA), or up to $\sim$ 10 years at $\sim$ 50 Kpc (e.g, SN1987A). In particular for the SN1987A case, the signature should have been detected by COMPTEL, if $^{22}$Na 
was produced by an H-ingestion event and CCSN explosion such as in the 25T-H model. 
However, its signature would have been extremely difficult to disentangle from the $^{56}$Co line at 1238 keV at these early stages due the instrument resolution. On the other hand, it will be possible to detect the predicted $^{22}$Na signature from new galactic CCSNe with future $\gamma$-ray telescopes like COSI and e-ASTROGRAM.  
\item We show that positron production from $^{22}$Na decay could significantly power the late CCSN light-curve, but it could also contribute to the observed positron annihilation rate in the galactic disk. We estimate its contribution to be small, in the order of a few percent, but current model uncertainties are too large to make any robust conclusion. 
\end{itemize}

Nucleosynthesis predictions from 1D models experiencing H-ingestion events are uncertain, and they should be taken only as a qualitative guidance. Multidimensional hydrodynamics models are needed to find out under which conditions H ingestion can take place at high metallicities, and to define the structure of He-burning layers following the ingestion of H. This would allow to quantify the local nucleosynthesis driven during the H ingestion and to constrain a realistic range for the amount of H left in the He-burning ashes \citep[][and references therein]{hoppe:24}. 
At the  moment, 3D simulations of convective H-He shell interactions in massive stars are not available for this evolutionary phase, and therefore the systematic uncertainties of any predictions regarding the 3D macro physics are unknown. In addition, even if these problems can be solved in the future, the necessary coupling to 1D evolution for time-scale bridging is another unsolved problem. The area of coupling of 3D macro physics events such as these to sufficiently large nuclear reaction networks is also still in its infancy \citep[][]{stephens:21}.  
Isotopic ratios in presolar grains may provide crucial benchmarks for such studies in the future. Furthermore, we have seen that several nuclear reaction rates are still uncertain in these nucleosynthesis regimes and more experiments and sensitivity studies are needed.

In this work, isotopic ratios measured in presolar stardust grains have again demonstrated how they 
can be used to define nucleosynthesis components and features that the next generation of models of massive stars and CCSNe will have to reproduce. The existence of the puzzling Ne-E(L) component in presolar graphites 
is another crucial piece of the puzzle to understand nucleosynthesis in CCSNe, and it should be taken into account within the context of CCSN models used to reproduce observations, from stellar archaeology and CCSN remnants, and for producing yields for galactic chemical evolution studies.

\begin{acknowledgments}
\textbf{The authors would like to thank the anonymous reviewer for the useful comments and suggestions provided for the manuscript.}
MP acknowledge the support to NuGrid from
the "Lendulet-2023" Program of the Hungarian Academy of Sciences (LP2023-10, Hungary), the ERC Synergy Grant Programme (Geoastronomy, grant agreement number 101166936, Germany), the ERC Consolidator Grant funding scheme (Project RADIOSTAR, G.A. n. 724560, Hungary), the ChETEC COST Action (CA16117), supported by the European Cooperation in Science and Technology, and the IReNA network supported by NSF AccelNet (Grant No. OISE-1927130). MP also thanks the support from NKFI via K-project 138031 (Hungary). The work by CLF was supported by the US Department of Energy through the Los Alamos National Laboratory. Los Alamos National Laboratory is operated by Triad National Security, LLC, for the National Nuclear Security Administration of U.S.\ Department of Energy (Contract No.\ 89233218CNA000001). SA acknowledges the support of NASA grant 80NSSC22K0360. AP acknowledges suppor from U.S. Department of Energy, Office of Science, Office of Nuclear Physics, under Award Number DE-SC0017799 and Contract Nos. DE-FG02-97ER41033. AML acknowledges the support of the Science and Technologies Facilities Council (STFC). RL thanks the support from the PRIN URKA Grant (Number prin$\_$2022rjlwhn). 
FH is supported by a Natural Sciences and Engineering Research Council of Canada (NSERC) Discovery Grant and acknowledges support from the NSERC award SAPPJ-797 2021-00032 $\emph{Nuclear physics of the dynamic origin of the elements}$.
We acknowledges support from the ChETEC-INFRA project funded by the European Union’s Horizon 2020 Research and Innovation programme (Grant Agreement No 101008324). We thanks access to {\sc viper}, the University of Hull HPC Facility. This research has used the Astrohub online virtual research environment (https://astrohub.uvic.ca), developed and operated by the Computational Stellar Astrophysics group (http://csa.phys.uvic.ca) at the University of Victoria and hosted on the Computed Canada Arbutus Cloud at the University of Victoria. This work benefited from interactions and workshops co-organized by The Center for Nuclear astrophysics Across Messengers (CeNAM) which is supported by the U.S. Department of Energy, Office of Science, Office of Nuclear Physics, under Award Number DE-SC0023128.
\end{acknowledgments}

%

\vspace{5mm}






\bibliography{references}{}
\bibliographystyle{aasjournal}



\end{document}